\documentclass[12pt,a4paper]{article}

\usepackage{indentfirst}
\usepackage{amsmath}
\usepackage{tabularx}
\usepackage{graphicx}
\usepackage{subfigure}
\usepackage{wrapfig}

\title{{\bf Electron dynamics at the initial stage of floating-sheath formation}}
\author{C. Lupu$^a$, D. D. Tskhakaya sr.$^{c,d}$, S. Kuhn$^c$, M. Cercek$^b$,\\ R. Schrittwieser$^e$, G. Popa$^a$}
\date{}

\begin{document}
\maketitle
\begin{center}
{\em $^a$ Plasma Physics Department, Faculty of Physics, Al. I. Cuza University, RO-700506 Iasi,
Romania,\\ $^b$J. Stefan Institute, University of Ljubljana, Jamova 39, SLO-1000 Ljubljana,
Slovenia,\\ $^c$ Association Euratom-OAW, Department of Theoretical Physics, University of
Innsbruck, A-6020 Innsbruck, Austria,\\ $^d$ Permanent address: Institute of Physics, Georgian
Academy of Sciences, 380077 Tbilisi, Georgia,\\ $^e$Association Euratom-OAW, Department of Ion
Physics, University of Innsbruck, A-6020 Innsbruck, Austria}
\end{center}

\begin{abstract}
The problem of sheath formation in front of a conductive planar plate inserted into the plasma is
formulated. Initially, the plate is assumed to be neutral.  It is shown that the charging-up
process of the plate is accompanied by the excitation of electron plasma waves.
\end{abstract}

\section{Introduction}
Investigations of sheath formation in front of a floating plate have hitherto been restricted to
fluid studies on the ion time scale [1]. By contrast, the response of the plasma in the very early
stages of sheath formation is not well known. In this paper, we present PIC simulations of the
plasma dynamics over just a few electron plasma periods after the beginning of the process. These
simulations have been performed by means of the BIT1 code [2], developed on the basis of the XPDP1
code from U. C. Berkeley [3].

A floating plate is placed in contact with a uniform, quasi-neutral plasma, which is assumed to be
infinitely extended on one side. Due to the higher thermal velocity of the electrons, the plate
starts charging up negatively, so that electrons are gradually repelled, ions are attracted, and a
positive-space-charge sheath begins to form. An electron plasma wave is observed the properties of
which strongly depend on the plasma characteristics (electron and ion temperatures, plasma
density, etc.).

Our PIC simulations are performed with different numerical set-ups and plasma characteristics. A
full set of simulation diagnostics is used to measure the properties of the electron waves.

\section{Simulation set-up}
We consider a one-dimensional system. The planar conducting plate and the (artificial) right-hand
boundary of the systems are placed at positions $x = 0$ and $x = L > 0$, respectively. The length
$L$ is to be chosen large enough for this system to reasonably approximate a semi-infinite plasma
($L >> \lambda_D$, with $\lambda_D$ the electron Debye length). In order to have adequate
resolution in space, the length of the grid cells has been selected as $\Delta x \approx
\lambda_D/2$.

\subsection{Initial and boundary conditions}
At the initial time $(t = 0)$ the electron and ion densities are equal $(n_{e0} = n_{i0} = n_0)$,
the distribution functions of both particle species are fully Maxwellian, and the electric
potential is zero $(V = 0)$ everywhere in the system, including the plate surface.

Throughout the entire simulation, the following boundary conditions are applied to the particles:
At the plate, all particles impinging are absorbed and no particles are injected into the plasma.
At the right-hand boundary, on the other hand, all particles impinging are absorbed but new
particles with half Maxwellian distribution functions are injected at a constant rate. The system
is floating, i.e., the sum of particle plus displacement currents equals zero. According to these
conditions we observe the following behavior.

In the unperturbed plasma region (i.e., for $x\rightarrow L$) the electron velocity distribution
function will not change appreciably (so that $E(L,t) = 0$), whereas at the plate it will acquire
a cut-off form. This is because the negative-velocity electrons are absorbed by the plate and
charge it negatively; during this process, the ions can be considered to be at rest. With
increasing negative surface charge, the negative potential drop in the region close to the plate
becomes higher and more and more electrons are reflected towards the plasma. After some time this
perturbation propagates into the system. The shape of the distribution function essentially
depends on the potential drop at the plate.

Due to the loss of particles by absorption at the plate, the total number of particles in the
system is dropping all the time. However, this aspect is not of great concern here because the
total loss of particles during the entire simulation presented is negligible.

\subsection{Simulation parameters}
In the following tables we present the parameters used for our simulation. The (electron and ion)
particle fluxes corresponding to the unperturbed plasma region are:
\begin{equation}
\Gamma_s=\frac{v_{th_s}}{\sqrt{2\pi}}\cdot n_{0s}\quad s=e,i
\end{equation}

These expressions are used to calculate the particle injection fluxes from the right-hand
boundary.

\vspace{0.1cm}
\begin{tabularx}{\linewidth}{|c|c|X|}
\hline
\hline
\multicolumn{3}{|c|}{\textbf{\textit{Plasma parameters}}}\\
\hline
\hline
\textbf{Parameter} & \textbf{Value} & \textbf{Remarks} \\
\hline
$T_e$ & $0.5\,eV$ &  \\
\hline
$T_i$ & $0.1\,eV$ &  \\
\hline
$n_{0i}=n_{0e}=n_0$ & $8.5\times10^{14}\,m^{-3}$ & at $t=0$\\
\hline
$v_{th_e}$ & $2.9649\times10^{5}\,m/s$ & \\
\hline
$v_{th_i}$ & $3.095\times10^{3}\,m/s$ & \\
\hline
$\Gamma_e$ & $1.006\times10^{20}\,m^{-2}s^{-1}$ & \\
\hline
$\Gamma_i$ & $1.05\times10^{18}\,m^{-2}s^{-1}$ & \\
\hline
$\omega_{pe}$ & $1.64\times10^{9}\,s^{-1}$& electron plasma frequency\\
\hline
$\omega_{pi}$ & $3.83\times10^{7}\,s^{-1}$& ion plasma frequency\\
\hline
$m_i$ & $1.67\times10^{-27}\,Kg$ & proton mass\\
\hline
$\lambda_D$ & $1.8\times10^{-4}\,m$& \\
\hline
\hline
\end{tabularx}

\vspace{0.1 cm}
\begin{tabularx}{\linewidth}{|c|c|X|}
\hline
\hline
\multicolumn{3}{|c|}{\textbf{\textit{Simulations parameters}}}\\
\hline
\hline
\textbf{Parameter} & \textbf{Value} & \textbf{Remarks} \\
\hline
$\Delta x$ & $8\times10^{-5}\,m$ & grid-cell length $\approx \lambda_D/2$\\
\hline
$L$ & $0.16\,m$ & system lenght\\
\hline
$S$ & $10^{-4}\,m^2$ & plate aria\\
\hline
$\Delta t$ & $5.398\times10^{-11}\,s$ & time step\\
\hline
$t_{tot}$ & $1.079\times10^{-7}\,s$ & total simulation time\\
\hline
\hline
\end{tabularx}

\section{Simulation results}
\begin{wrapfigure}{r}{0.5\textwidth}
\centering
\includegraphics[width=0.5\textwidth]{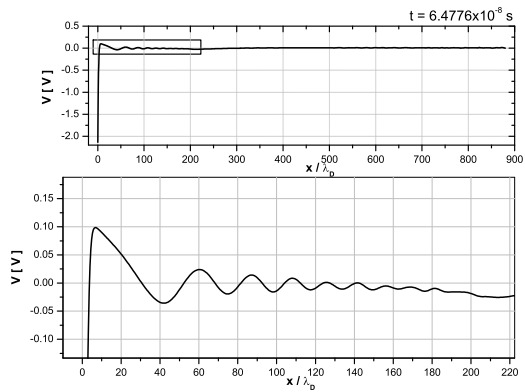}
\caption{{\it Potential profile in the perturbation region at $t=6.4776\times 10^{-8}\,s$ }}
\label{pot_strat_sursa}
\end{wrapfigure}
\begin{figure}[!h]
\centering
\includegraphics[width=0.9\textwidth]{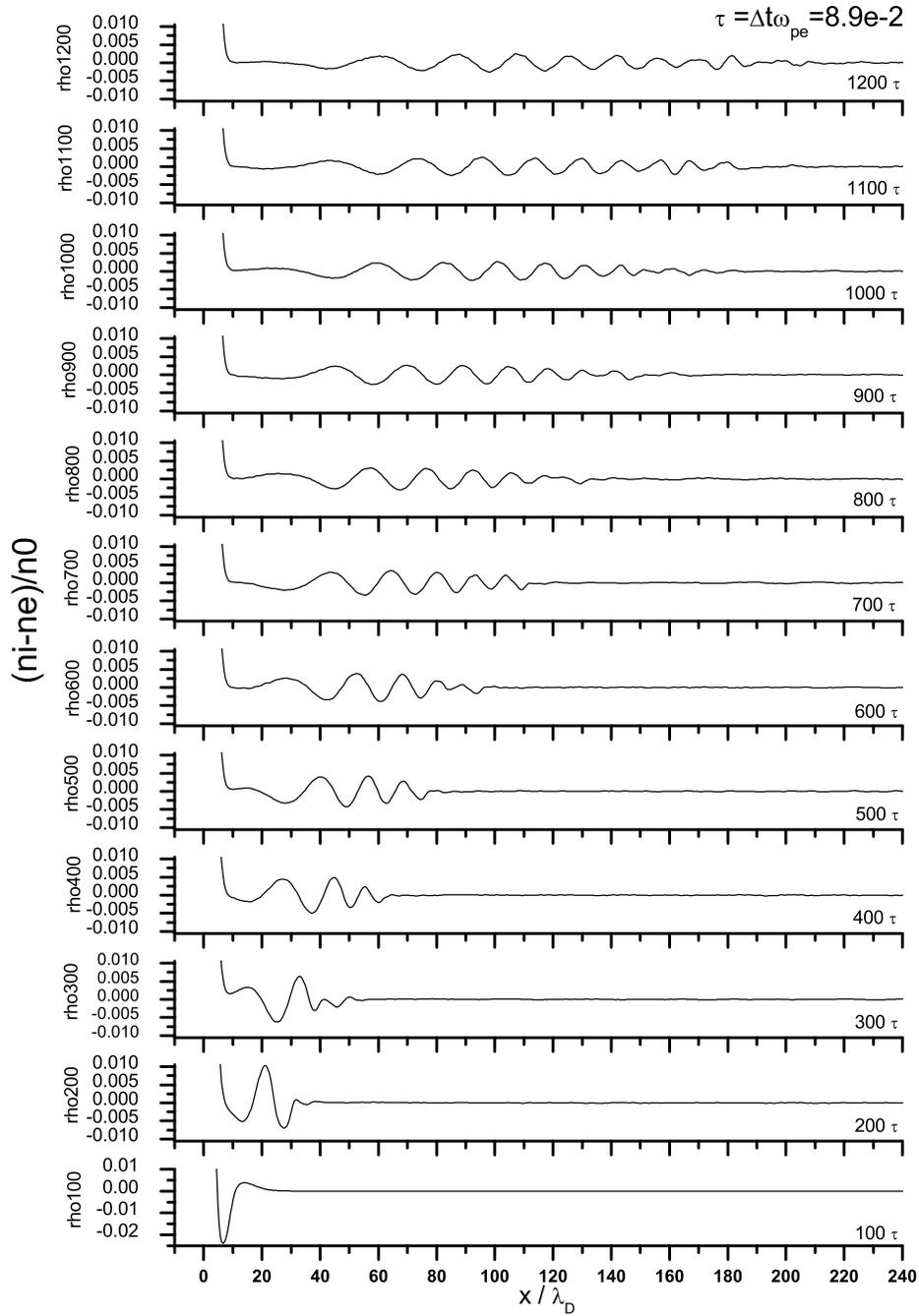}
\caption{\textit{Electron waves evolution}}
\label{evol_rho}
\end{figure}
\begin{wrapfigure}{r}{0.5\textwidth}
\centering
\includegraphics[width=0.5\textwidth]{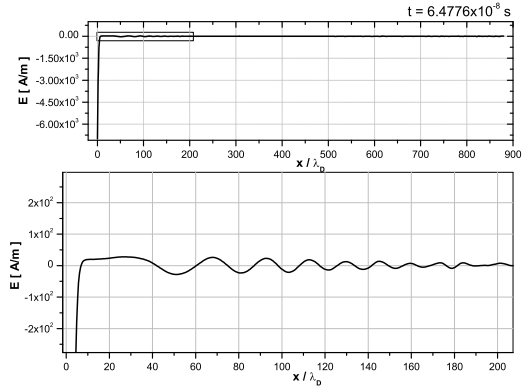}
\caption{\textit{Electric field profile (perturbation region); $T_e=0.5eV$}}
\label{cimp_strat}
\end{wrapfigure}

Figure \ref{pot_strat_sursa} shows the potential profile close to the plate at $t = 6.477\times
10^{-8}$ s. The potential drop at the beginning of the sheath evolution is monotonic in space.
After quick acquisition of negative charge, the plate repels the electrons in the form of a pulse
leaving behind a positive-space charge region. As a result, the potential close to the plate
becomes slightly positive. In front of this region, the negative space charge produced by the
primary-pulse electrons leads to a potential minimum ("virtual cathode"), which gradually reflects
more and more slower electrons back into the plasma. These latter electrons spend a long time in
the region of the virtual cathode and hence deepen its potential further. According to Figures.
\ref{evol_rho} and \ref{cimp_strat}, this first potential perturbation (consisting of a potential
hill and a potential well) propagates into the unperturbed plasma, with additional similar wave
structures forming behind it.

\begin{figure}[!h]
\centering
\includegraphics[width=0.7\textwidth]{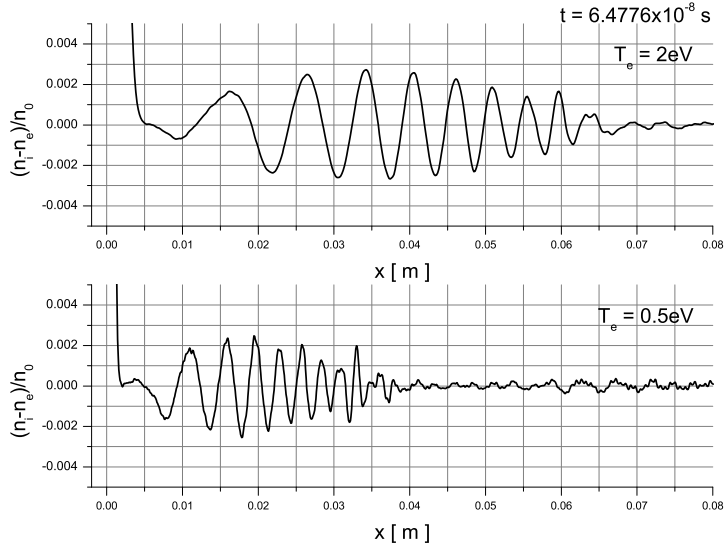}
\caption{\textit{Electron waves in plasma with different temperatures for electrons $T_e=2eV$ and
$T_e=0.5eV$ }}
\label{comparare}
\end{figure}

To verify that these waves are physical and not just due to numerical effects, we have performed
other simulations with different parameters. In particular, we concentrated on the electron
temperature. We know that the Debye length is proportional to the square root of the electron
temperature. Hence, if we increase the temperature by a factor of four, the Debye length must
increase by a factor of two. Since, in addition, there is a relation between the wavelength of the
electron waves and the Debye length, the variation of the electron temperature should also have an
effect on the wavelength. This is clearly illustrated in \\Figure \ref{comparare}, where the
wavelength is seen to increase with the square root of the electron temperature.

\section{Summary and conclusions}
This work represents the beginning of a self-consistent kinetic study of sheath formation, taking
into account both electron and ion dynamics. Here, during the short simulation time considered,
the ions are practically immobile, and only the electrons take part in the process. In the next
step, the effect of ion dynamics on sheath formation will be considered as well.

\section*{Acknowledgements}

This work was supported by the Austrian Science Fund (FWF) Projects P15013-N08 and P16807-N08,
CEEPUS Network A103, and Erasmus/Socrates grant 2004-2005.

\section*{References}
{\bf [1]} J.W. Cipolla, Jr., and M. B. Silevitch, On the temporal development of a plasma sheath,
J. Plasma Phys. 25, 373-89 (Jun 1981)

{\bf[2]} D. Tskhakaya and S. Kuhn, Effect of EB drift on the plasma flow at the magnetic
presheath entrance, Contrib. Plasma Phys. 42, 302 (2002).

{\bf [3]} J. P. Verboncoeur, M. V. Alves, V. Vahedi, and C. K. Birdsall, Simultane­ous potential
and circuit solution for 1D bounded plasma particle simulation codes, J. Comput. Phys. 104 (2),
321 (1993). Abstract Submittal Form


\end{document}